# Laboratoire de l'Accélérateur Linéaire

# Physics from Polarized *ep* Collisions at HERA

Zhiqing Zhang

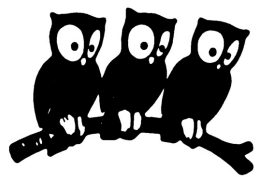

# Physics From Polarized *ep* Collisions at HERA

Zhiqing Zhang

*Laboratoire de l'Accélérateur Linéaire, IN2P3-CNRS et Université de Paris-Sud XI, Bât. 200, BP 34, 91898 Orsay Cedex, France (zhangzq@lal.in2p3.fr)*

**Abstract.** Physics from the polarized electron or positron beam colliding with the unpolarized proton beam at HERA-II as well as from the unpolarized *ep* collisions at HERA-I is discussed. Results shown include (i) inclusive neutral and charged current cross sections and structure functions measurements at HERA-I, (ii) the impact of these measurements on parton density functions, on electroweak parameters and searches for new physics and (iii) the first results at HERA-II and its future prospects.

**Keywords:** Deep inelastic lepton-proton scattering, neutral and charged current cross sections, proton structure functions, parton distribution functions, electroweak parameters, polarized lepton beams.
**PACS:** 13.60.-r, 12.38.-t, 12.15.Ji.

## INTRODUCTION

The HERA electron- or positron-proton collider is a unique facility for studying neutral and charged current (NC, CC) processes in deep inelastic scattering (DIS) interactions in an extended kinematic region compared to that covered previously by fixed target experiments. HERA-I operation started in 1992 and ended in 2000. Earlier measurements at HERA-I from H1 and ZEUS revealed the strong rise of the structure function (SF) $F_2$ towards small $x$. The more precise measurement of NC and CC cross sections is based on three recent data samples of $e^+p$ taken in years from 1994 to 1997 at a center-of-mass (CM) energy of 301 GeV, $e^-p$ in 1998-1999 and $e^+p$ in 1999-2000 at 319 GeV. The corresponding integrated luminosities per experiment of the three data samples are around 40 pb$^{-1}$, 15 pb$^{-1}$ and 65 pb$^{-1}$, respectively. The HERA-I data have driven the determination of parton density functions (PDFs) in various global fits since a decade. The universal nature of the PDFs makes precise determination essential in order to provide reliable predictions for Standard Model (SM) processes and for searches for new physics at future hadron-hadron colliders such as the LHC. Since 2002, HERA has entered its second phase of operation (HERA-II) and is being run in higher luminosity mode providing in addition longitudinally polarized positron and electron beams for H1 and ZEUS. The improved luminosity is due to a finer beam focusing around the interaction points and to larger beam currents.

In the following, I shall start with the general neutral and charged current cross section formulae showing how the cross sections are modified when the electron beam is polarized. I will continue to briefly review the precise cross section and SF data measured at HERA-I and their impact on the determination of PDFs, electroweak (EW) parameters and on searches for new physics. I then show first results obtained using the HERA-II data before discussing future prospects.

## NEUTRAL AND CHARGED CURRENT CROSS SECTIONS

Both NC and CC processes can be produced at HERA. The NC interaction is dominated by $\gamma$ exchange in the $t$ channel at low $Q^2$ and receives additional contributions from $Z$ exchange and $\gamma Z$ interference at high $Q^2$. The NC cross section may be expressed in terms of three SFs $\tilde{F}_2$, $\tilde{F}_L$ and $x\tilde{F}_3$ as

$$\frac{d^2\sigma_{NC}(e^\pm p)}{dxdQ^2} = \frac{2\pi\alpha^2}{xQ^4}\left[Y_+\tilde{F}_2 - y^2\tilde{F}_L \mp Y_-x\tilde{F}_3\right] \tag{1}$$

where $\alpha$ is the fine structure constant and $Y_\pm = 1 \pm (1-y)^2$ with the inelasticity variable $y$ being connected to other kinematic variables and the CM energy squared $s$: $y=Q^2/xs$. The generalized SFs $\tilde{F}_2$ and $x\tilde{F}_3$ can be further decomposed as

$$\tilde{F}_2 = F_2 - (v_e - P_e a_e)\kappa_Z F_2^{\gamma Z} + (v_e^2 + a_e^2 - 2P_e v_e a_e)\kappa_Z^2 F_2^Z$$
$$x\tilde{F}_3 = -(a_e - P_e v_e)\kappa_Z xF_3^{\gamma Z} + \left[2v_e a_e - P_e(v_e^2 + a_e^2)\right]\kappa_Z^2 xF_3^Z \tag{2}$$

with $v_e$ and $a_e$ being vector and axial-vector weak couplings of electron to $Z$, $P_e$ the longitudinal polarization of the electron beam and

$$\kappa_Z^{-1} = \frac{Q^2 + M_Z^2}{Q^2}\frac{2\sqrt{2}\pi\alpha}{G_F M_Z^2} \tag{3}$$

in the modified on-mass-shell scheme, in which all EW parameters can be expressed in terms of $\alpha$, the Fermi coupling constant $G_F$ and the $Z$ boson mass $M_Z$. The SF $F_2$ originates from photon exchange only. The SFs $F_2^Z$ and $xF_3^Z$ are the contributions to $\tilde{F}_2$ and $x\tilde{F}_3$ from $Z$ exchange and the SFs $F_2^{\gamma Z}$ and $xF_3^{\gamma Z}$ are the contributions from $\gamma Z$ interference. The longitudinal SF $\tilde{F}_L$ may be decomposed in a manner similar to $\tilde{F}_2$.

In the quark parton model (QPM), the SF $\tilde{F}_L \equiv 0$, the SFs $F_2$, $F_2^{\gamma Z}$ and $F_2^Z$ are related to the sum of the quark and anti-quark distributions

$$\left[F_2, F_2^{\gamma Z}, F_2^Z\right] = x\sum_q \left[e_q^2, 2e_q v_q, v_q^2 + a_q^2\right](q+\bar{q}) \tag{4}$$

and the SFs $xF_3^{\gamma Z}$ and $xF_3^Z$ to their difference

$$\left[xF_3^{\gamma Z}, xF_3^Z\right] = 2x\sum_q \left[e_q a_q, v_q a_q\right](q-\bar{q}) \tag{5}$$

where $v_q$ and $a_q$ are weak neutral current couplings of quark $q$ to the Z boson.

The CC interaction takes place by $W$ exchange only and its cross section may be expressed in a similar way as

$$\frac{d^2\sigma_{CC}(e^\pm p)}{dxdQ^2} = (1\pm P_e)\frac{G_F^2}{4\pi x}\left[\frac{M_W^2}{M_W^2 + Q^2}\right]^2 \left[Y_+ W_2 - y^2 W_L \mp Y_- xW_3\right] \tag{6}$$

with $M_W$ being the mass of the W boson.

# HERA-I ACHIEVEMENTS

## Measurements of Structure Functions and Neutral and Charged Current Cross Sections

The dominant SF $\tilde{F}_2$, which is sensitive to all quark species, has been determined from the measured NC cross sections over more than 4 orders of magnitude both in $x$ and in $Q^2$. The best precision reaches an accuracy down to 2% (Fig.1) [1,2]. The contribution of the SF $\tilde{F}_L$, which is sensitive to higher order gluon radiation processes thereby providing valuable information on the gluon content of the proton, is sizable only at high $y$ and can only be determined indirectly for the moment [1]. The SF $x\tilde{F}_3$, which is sensitive to valence quarks at high $x$, is non-negligible at high $Q^2$ and is determined [3-5] using both $e^+p$ and $e^-p$ data samples with so far limited precision.

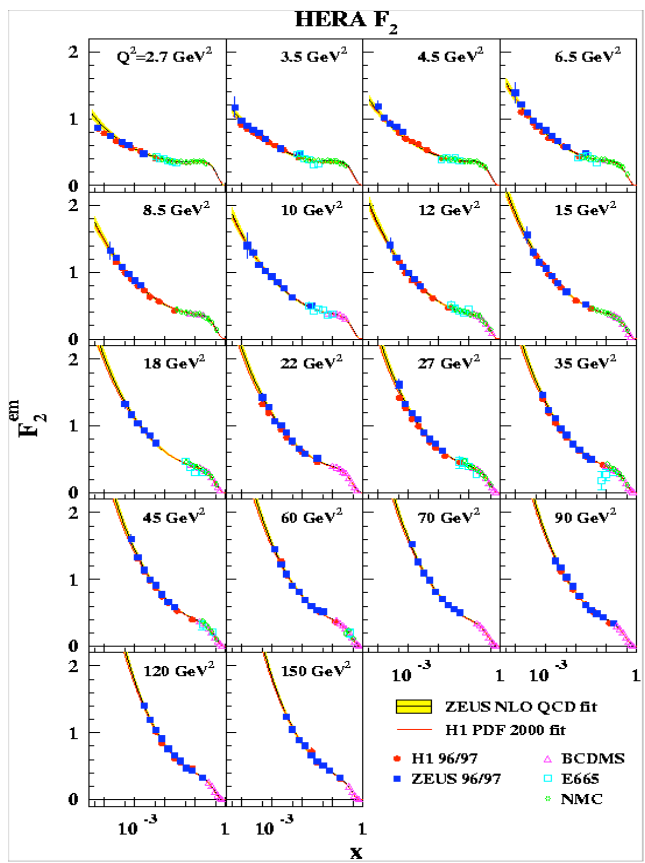

**FIGURE 1.** The pure electromagnetic ($\gamma$ exchange) structure function $F_2^{em}$ in $\tilde{F}_2$ measured by HERA experiments H1 and ZEUS and by fixed target experiments BCDMS, E665 and NMC. The curves show the corresponding expectations from ZEUS next-leading-order (NLO) QCD fit and H1 PDF 2000 fit (see text below).

While the NC SF $\tilde{F}_2$ is essentially flavor blind, the CC cross sections are flavor dependent. In particular the measured CC $e^+p$ cross section data [3,4,6,7] can be used to uniquely constrain the down-type quark density (mainly $d$ quark) free from any nuclear corrections. This is illustrated in Fig.2 where the measured $e^-p$ CC cross sections are found to be larger than the corresponding $e^+p$ CC cross section. The difference is due to the larger up-type quark density than the down-type quark density at high $x$ and more favorable helicity factor in the $e^-p$ collisions. In fact removing the kinematic factors from Eqn.(2), a reduced CC cross sections ($\tilde{\sigma}$) can be defined as

$$\tilde{\sigma} = \frac{d^2\sigma_{CC}(e^\pm p)}{dxdQ^2}\frac{4\pi x}{G_F^2}\left[\frac{M_W^2+Q^2}{M_W^2}\right]^2 = \left[Y_+W_2 - y^2W_L \mp Y_-xW_3\right]. \quad (7)$$

At leading order in the QPM model it can be expressed in terms of up- and down-type quark densities as $\tilde{\sigma}_{CC}^{e^-p} = (u+c)+(1-y)^2(\bar{d}+\bar{s})$ and $\tilde{\sigma}_{CC}^{e^+p} = (\bar{u}+\bar{c})+(1-y)^2(d+s)$.

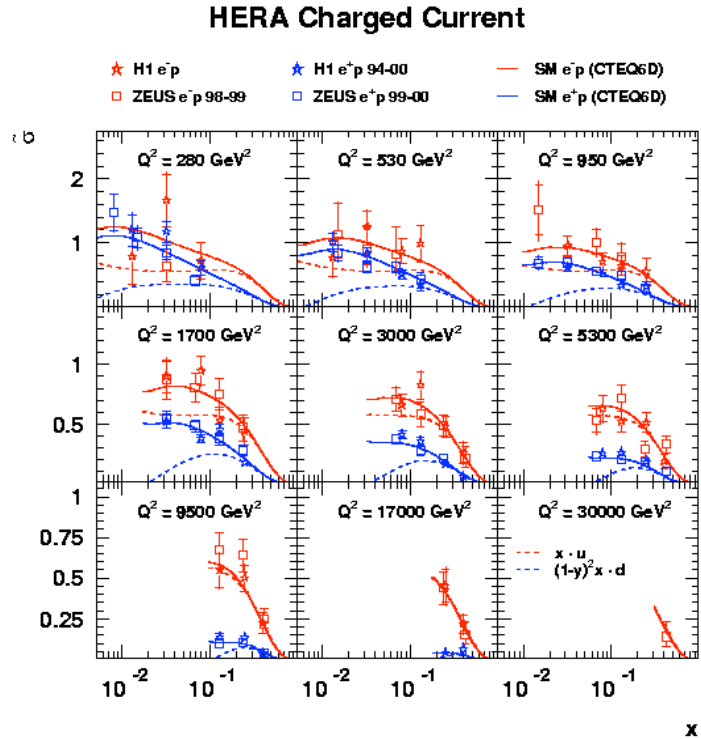

**Figure 2.** Measured charged current $e^\pm p$ cross sections by H1 and ZEUS in comparison with the corresponding Standard Model (SM) expectations based on the CTEQ6D parameterization. The dashed curves show the dominant contribution at high $x$.

## Impact of the HERA-I Data on Patron Distribution Functions

HERA-I data have been extensively used by the MRST and CTEQ groups [8,9] in their global fits to determine the PDFs in the last decade with steadily increasing precision. The impact of the HERA-I data is best seen in QCD analyses using these data only. Indeed, the fit performed by H1 [1], based on all $e^+p$ and $e^-p$ NC and CC cross section data at low and high $Q^2$, was used to determine 5 PDF sets $xf$: $xg$, $xU=x(u+c)$, $xD=x(d+s)$, $x\bar{U} = x(\bar{u}+\bar{c})$ and $x\bar{D} = x(\bar{d}+\bar{s})$. The results of the fit are shown in Fig.3 compared to PDFs determined in a similar analysis from ZEUS [10] using, in addition to cross section data, jet data in order to improve the constraint on the gluon. The agreement between H1 and ZEUS is fairly good and the visible difference in e.g. the gluon distribution may be understood due to different functional forms, heavy flavor treatments and constraints of the data sets.

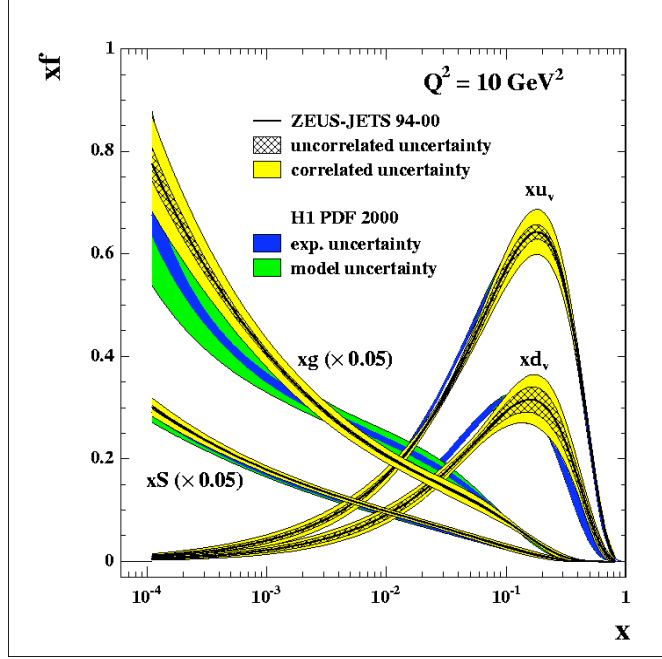

**FIGURE 3.** A comparison of PDFs determined by H1 (H1 PDF 2000) based on inclusive NC and CC cross section data and by ZEUS (ZEUS-JETS) using jet data in addition to the cross section data. The gluon (*xg*) and sea (*xS*) densities are scaled by a factor 20.

## Determination of Electroweak Parameters at HERA-I

The inclusive NC and CC cross sections are not only primary source of PDFs but have also sensitivity to EW parameters. Indeed, the NC cross section depends on up- and down-type quark couplings to the $Z^0$ boson, $a_q$ and $v_q$ ($q=u, d$), via SFs (Eqns.(4,5)). Equation (6) shows that the shape of the CC cross section as a function of $Q^2$ is controlled by the propagator mass ($M_{prop}$) of the $W$ boson whereby it was previously determined by both collaborations (see e.g. [4,6,7]). In a recent combined EW-PDF analysis by H1 [11,12], using the full HERA-I data and taking properly into account the correlation between $M_{prop}$ and PDFs, an improved measurement of $M_{prop}$ has been obtained yielding $M_{prop} = 82.87 \pm 1.82_{exp}{}^{+0.30}_{-0.16}|_{model}$ GeV where the first error is experimental and the second error covers model uncertainties [11,12]. This measurement based on the virtual $W$ boson exchanged in $t$ channel in the space-like regime is independent of the other direct measurements performed using real $W$ bosons produced singly at the Tevatron and in pairs at LEP-2. Within the SM, the Fermi coupling constant $G_F$, which determines the normalization of the CC cross section, is connected with the $W$ boson mass $M_W$ through the relation

$$G_F = \frac{\pi\alpha}{\sqrt{2}M_W^2\left(1 - \frac{M_W^2}{M_Z^2}\right)} \frac{1}{1 - \Delta r} \quad (8)$$

where the term $\Delta r$ contains one-loop and leading higher-order EW radiative corrections including quadratic (logarithmic) dependence on the top quark mass $m_t$ (the Higgs mass $M_H$). Using the SM relation (8), a similar combined EW-PDF fit gives $M_W = 80.786 \pm 0.205^{+0.048}_{exp\,-0.029}|_{model} \pm 0.025_{\delta m_t} - 0.084_{\delta M_H} \pm 0.033_{\delta(\Delta r)}$ GeV where the measured central value corresponds to using the world averaged values of $M_Z$=91.1876±0.0021GeV, $m_t$=178±4.3GeV and a Higgs mass of 120GeV. The uncertainty on $M_Z$ has a negligible error on $M_W$ whereas the uncertainty on $m_t$ gives rise to the third quoted error on $M_W$. Varying $M_H$ from 120GeV to 300GeV results in the fourth error. The last error is due to higher order radiative correction uncertainties.

The light quark couplings to $Z^0$ have also been determined in another combined fit [12]. The results are shown in Fig.4 and compared with those obtained by CDF [13] and LEP [14]. The H1 precision is comparable with that of CDF. The combined LEP results are more precise but have ambiguities in sign and between $a_q$ and $v_q$.

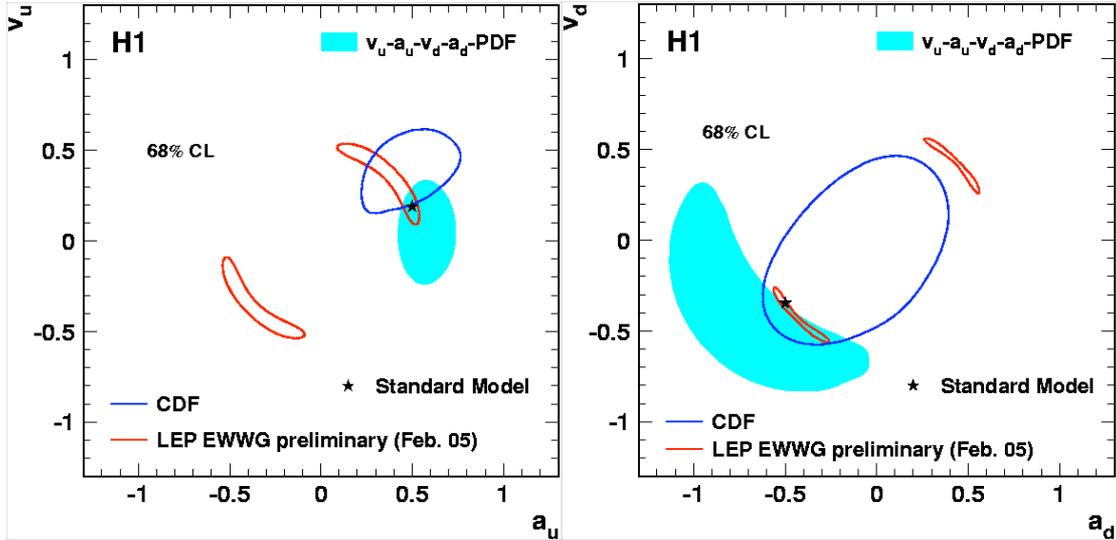

**FIGURE 4.** H1 results (shaded area) at 68% confidence level (CL) on the weak neutral couplings of $u$ quark (left) and $d$ quark (right) to $Z^0$ in comparison with similar results from CDF (blue curves) and LEP (red curves).

## Searches for New Physics

The inclusive NC and CC data have been furthermore used in searching for new physics. Indeed, the HERA $ep$ collider is ideally suited for resonance production of leptoquarks LQs which have decay final states similar to that of the SM NC and CC processes but differ in angular distributions. No evidence for LQ production was so far observed and stringent upper limits on LQ couplings to standard fermions as functions of LQ masses were derived [15,16]. Figure 5 shows the limit set by H1 for an example scalar LQ ($\tilde{S}_{1/2,L}$) in comparison with the direct search limit from D0 [19] and the best indirect search limit at LEP from OPAL [20].

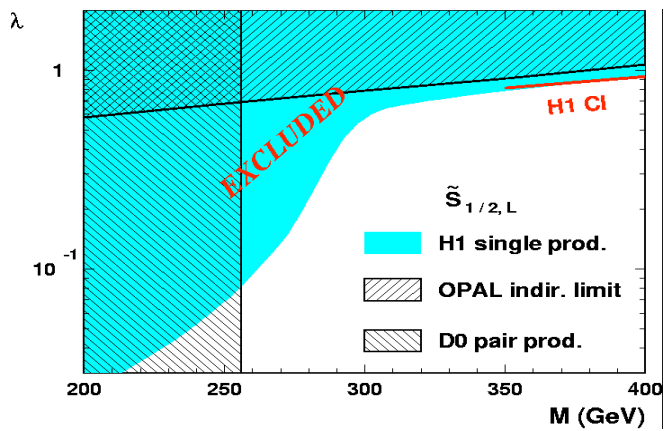

**Figure 5.** Exclusion limit at 95% confidence level on the coupling λ as a function of LQ mass $M$ for $\tilde{S}_{1/2,L}$ in the framework of Buchmüller-Rückl-Wyler model [17]. The H1 limit is compared with a direct limit from D0, an indirect limit from OPAL and an H1 limit from a contact interaction (CI) analysis [18].

# HERA-II AND FUTURE PROSPECTS

## HERA-II Status and First Results

In order to increase the luminosity delivered, the HERA accelerator was upgraded in 2001. During the first two years running after the upgrade, however, the experiments suffered large backgrounds. The problem has now been solved and an increase in peak luminosity around a factor 3 was achieved in the running since 2004. In addition, two pairs of spin rotators have been installed near the H1 and ZEUS interaction points providing for the first time longitudinally polarized positron and electron beams to these experiments.

HERA-II has delivered several independent data samples for both $e^+p$ and $e^-p$ collisions with positive and negative polarizations. They have been used by both experiments to measure the integrated CC cross sections. Together with the CC cross section measured [3,7] previously at HERA-I with unpolarized data, the dependence of the CC cross section on polarization is now available (Fig.5). In the SM, the cross section has a linear dependence as a function of polarization as represented by the straight lines based on the expectations from MRST. Due to the absence of right-handed $W$ bosons in the SM, the CC $e^+p$ ($e^-p$) cross section at $P_e=-1$ ($P_e=1$) is predicted to vanish. The measurements are consistent with the expectations. Future measurements with improved precision will provide a stringent constraint on the contribution of the right-handed *W* boson and its mass, independent of that from the Tevatron. The HERA-II NC data have also been analyzed and the small polarization effect expected at high $Q^2$ on the NC cross sections have not yet been established experimentally [21].

There have been a number of interesting observations obtained based on the HERA-I data. One example is the excess with respect to the SM expectation of events having an isolated charged lepton and large missing transverse momentum in the final state [22]. All presently available HERA-II data have been analyzed. The current situation is summarized in Table 1. It shows that as far as the electron channel is concerned, H1 still observes an overall excess at higher missing transverse momentum values $P_{T,X}$>25GeV whereas the similar analysis from ZEUS does not confirm the H1 observation. On the other hand, ZEUS does observe an excess in the tau channel based on HERA-I data [23]. One eagerly awaits more data from HERA-II to check whether this is due to a statistical fluctuation or a sign of new physics.

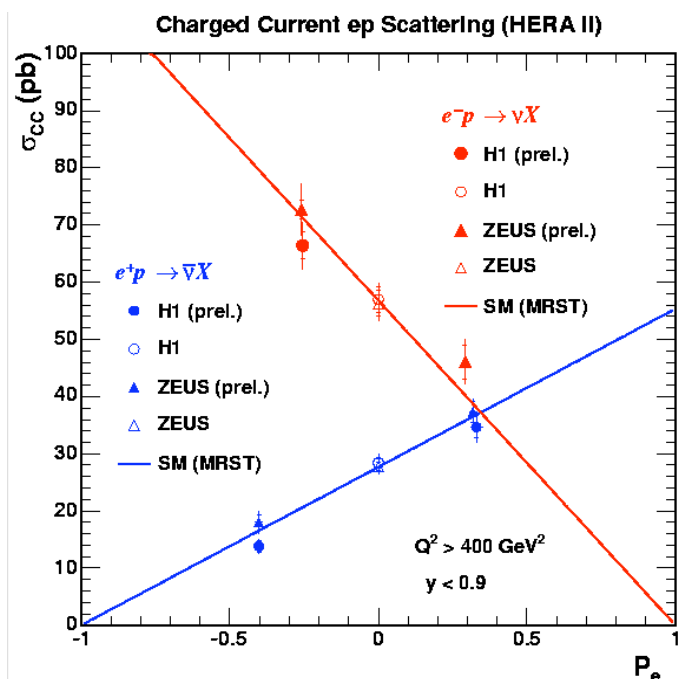

**FIGURE 5.** The dependence on polarization of the integrated CC cross sections shown for $Q^2$>400GeV$^2$ and *y*<0.9. The straight lines corresponds to the expectation from MRST.

**TABLE 1.** Present status of the analysis of events with an isolated electron candidate and large transverse momentum ($P_{T,X}$). The first column indicates the data set, the collision mode and the integrated luminosity in parentheses. In the second column, the number of observed events and the corresponding SM expectations are given with the percentage number in parentheses showing the dominant contribution from the single *W* production. The last column shows the same information as the second column but at higher $P_{T,X}$ values.

| Isolated *e* candidates | 12<$P_{T,X}$<25 (GeV) | $P_{T,X}$>25 (GeV) |
|---|---|---|
| H1 1994-2000 $e^+p$ (104.7 pb$^{-1}$) | 1/1.96±0.27(74%) | 4/1.48±0.25(86%) |
| H1 (prel.) 1994-2005 $e^{\pm}p$ (211 pb$^{-1}$) | − | 11/3.2±0.6(77%) |
| ZEUS (prel.) 1999-2000 $e^+p$ (66 pb$^{-1}$) | 1/1.04±0.11(57%) | 1/0.92±0.09(79%) |
| ZEUS (prel.) 2003-2004 $e^{\pm}p$ (40 pb$^{-1}$) | 0/0.46±0.10(64%) | 0/0.58±0.09(76%) |

# Future Prospects

The HERA-II operation will end in 2007 according to the current planning. In the present envisaged running scenario HERA will continue to operate with an electron beam till some time in the beginning of 2006 and will switch back to positron-proton collisions for the rest of HERA-II. In this scenario, an integrated luminosity of 0.7 fb$^{-1}$ per experiment is expected. This will be a significant gain with respect to the total data taken in HERA-I. Such a large data sample will improve even further our knowledge of the PDFs, in particular at high $x$ and $Q^2$, the precision of which is currently statistically limited.

Indeed, precision NC cross sections at high $Q^2$ will improve the determination of $x\tilde{F}_3$ which is sensitive to $2u_v+d_v$ with $u_v$ and $d_v$ being the valence quark densities of $u$ and $d$ quarks. Similarly, precision $e^+p$ CC cross sections at high $Q^2$ will allow the $d$ quark density to be constrained better at high $x$. Furthermore, from a polarization asymmetry in NC cross sections, one will determine the SF $F_2^{\gamma Z}$ which can be expressed in terms of the $d/u$ ratio. All these data will thus hopefully clarify the controversial value of $d/u$ at $x$=1 [24].

The longitudinal polarization of electron and positron beams also provides new measurement possibilities. One example is the additional constraint on the vector coupling $v_q$ of light quarks to $Z$. According to Eqn.(2), when the polarization is non-zero, the second term $F_2^{\gamma Z}$ of $\tilde{F}_2$ is no longer negligible thus giving access to $v_q$ (Eqn.(4)).

Apart from the high luminosity run at largest possible beam energies, a direct measurement of $F_L$ will need data at different CM energies. Such a measurement is important to verify the gluon density that is currently constrained mainly in an indirect way from the scaling violation of the SF data. A lower proton energy running would also help to access high $x$ region at low $Q^2$.

Beyond HERA-II, there is strong interest [25] in the community to continue the HERA physics programme and to fully exploit its unique potential. In particular it is expected that by replacing the proton beam with a heavy nuclear beam, one can study a new QCD regime in which the gluon parton density becomes so high that saturation would take place. This however needs funding and solutions for a new injector replacing PETRA, the actual injector, which will be rebuilt to be used for XFEL (X-ray Free Electron Laser) from 2007 on.

# ACKNOWLEDGMENTS

The author wishes to thank his colleagues in H1 and ZEUS for some of the results shown in this contribution.